# Enhanced Evanescent Field Spectroscopy at Waveguide Surfaces using High Index Nano and Near-Nano Layers

**John Canning,**[1,2,*] **Whayne Padden,**[1] **Danijel Boskovic,**[1,2] **Masood Naqshbandi,**[1,2] **Lorenzo Costanzo,**[2] **Hank de Bruyn,**[2] **Tze H. Sum,**[2] **and Maxwell J. Crossley,**[2]

[1]*Interdisciplinary Photonics Laboratories (iPL), School of Chemistry, 222 Madsen Building F09, University of Sydney, Sydney NSW 2006, Australia*
[2]*School of Chemistry, Building F08, University of Sydney, Sydney NSW 2006, Australia*
[*]*john.canning@sydney.edu.au*

**Abstract:** We propose and demonstrate, through simulation and experiment, how the interaction of an optical field within a waveguide designed for chemical sensing and, more generally, evanescent field spectroscopy can be enhanced substantially by strategic deposition of high index surface layers. These layers draw out the optical field in the vicinity of probing and take advantage of field localisation through optical impedance matching. Localisation of the evanescent field to the inner layer in turn is accompanied by whispering gallery modes within the channels of a structured cylindrical waveguide, further enhancing sensitivity. A novel demonstration based on self-assembled layers made up of $TiO_2$ within a structured optical fibre is demonstrated, using a simple porphyrin as the spectroscopic probe. This technique offers optimisation of the limitations imposed on practical waveguide sensors that are highly sensitive but nearly always at the expense of low loss. The principles have potential ramifications for nanophotonics more generally and these are discussed.

## 1. Introduction

Evanescent field spectroscopy (EFS) using optical waveguides, both conventional [1] and structured [2] is an extremely powerful tool, both in terms of chemical sensor applications and as well as fundamental molecular studies of layers and films that circumvents the need for high finesse resonant spectroscopy, including cavity ring down spectroscopy, where temporal responses and lifetimes may need special consideration, for ultra high sensitivity. EFS using conventional fibres has some drawbacks since interaction with the outside usually requires exploitation of the cladding mode evanescent fields; coupling to and from the core travelling mode is, for example, achieved via phase matching through long period gratings [1]. Whilst there are other configurations, structured fibres have channels where the core traveling modes directly overlap with the channels and therefore core mode evanescent fields are exploited. By having arbitrary long interaction lengths using structured optical fibres, for example, unprecedented direct sensitivity is, in principle, possible. Indeed, we recently reported the observation of a near IR band associated with charge transfer between dichloro[5,10,15,20-tetra(heptyl)porphyrinato]tin(IV) [Cl−Sn(THP)-Cl] a porphyrin molecule and the inner silica surface of the channels of a structured optical fibre over 90 cm in length [2], previously only postulated. For many applications, particularly in distributed chemical and biochemical, but also in novel optoelectronic devices that rely on overlap with the evanescent field, these lengths become impractical. Solutions involve ring waveguide configurations that are essentially integrated resonant spectrometers. A popular research alternative to EFS involves using bandgap waveguides, such as the Bragg, or Fresnel, fibre [3-5] and its structured variants [5-7], which have confined optical fields propagating within the guiding hole of the waveguide containing the sample under test – however, these bandgap waveguides can be extremely sensitive to perturbations, such as temperature and strain, which shift the bands and affect interactions. They are also difficult to couple to standard telecommunications fibres, making remote sensing challenging. In any case, for many applications it is the field at the interface, particularly for selective surfaces, which is important and EFS remains the preferred interrogation tool. Similar surface plasmon resonances demonstrated using photonic bandgap waveguides [7] can in principle also be integrated into conventional structured optical fibres.

Alternative variants to optical fibres have reflected a strong push for more compact waveguides with substantially larger evanescent field interactions such as silicon-air slot waveguides [8,9] and nanowires [10-12]. These systems are often characterised by materials with higher refractive indices than silica or rely on extraordinarily small waveguide cross-sections to enhance the evanescent field intensity. As a consequence they are currently associated with high losses, often limited in fabrication to very short lengths, have poor reproducibility or have a major issue in terms of coupling light into and out of them – the balance between high sensitivity, low loss and practical implementation becomes a clear limiting factor and is a major research front. As well, for many practical applications remote operation is increasingly important – chemical sensing in the oil industry may require lead lengths of kilometers in order to perform highly accurate spectroscopy down an oil bore, for example. The retention of low propagation and low coupling loss, and stable and robust attachment, therefore, points back to optical material platforms such as $SiO_2$ whether as fibre or in a form which can be readily and robustly integrated with zero or negligible loss. Yet it is clear there are limitations with regards to sensitivity at the measurement head, or at the point of sensing, of such a waveguide – typically, low loss, standard telecommunications fibre compatible structured waveguides have only a few % of light in the evanescent field. Practical size of such heads also determines multi-sensor capability – drawing down to smaller tapers as a solution, typically suggested to offset such issues, requires longer lengths in order to ensure adiabatic transformation of the optical mode and low loss. It also increases fragility.

The use of optical impedance matching (similar to electromagnetic impedance matching in microstrip lines) to enable localised evanescent optical fields to reach high intensities for sensing work is a relatively recent proposal not only with moderate index silicon slot waveguides [8,9] but within holes inside low index silica structured fibres such as Fresnel fibres and photonics crystal fibres [13,14]. Interestingly, direct experimental observation of edge localisation has not been reported. Generally, the smaller the hole the substantially larger the field within the hole – for silica this is ideally sub-wavelength. For many applications, however, such as chemical and bio-sensing, holes have to be sufficiently large to overcome rate limiting steps such as diffusion (Brownian or Fickian), occasionally kinetic related impediments (electro-osmosis and electrophoresis when charged particles are involved) and potentially even van der Waals forces when holes become very small. Below about 1μm, these effects can prevent any intake of a sample, for example. On the other hand, there is clearly a potential advantage to exploiting the accumulation of evanescent field, improving sensitivity and efficiency. In this paper, we propose an alternative approach avoiding sub wavelength holes. It is based on enhancing the sensitivity of robust structured waveguides generally by depositing a higher index layer to draw out the guided optical light to enhance the evanescent field interactions but simultaneously minimise additional confinement losses. This approach allows the retention of platform backbone technology based on silica. It also has much broader implications for nanophotonics and nanotechnology generally.

## 2. The concept and simulation

As a demonstrator, we consider the waveguide configuration often used in optical sensors – the so-called structured optical fibre, such as a photonic crystal silica fibre, where propagation is determined largely by effective step-index total internal reflection [6]. A schematic of a simple two ring structure is shown in Figure 1 – for practical purposes the bulk of the optical field is confined by the first ring of holes. Propagation and the waveguide mode field distribution, including the evanescent field within the holes, are evaluated numerically using a full vectorial algorithm for 2-D structures which has been successfully used to design various structured and diffractive fibres [http://code.google.com/p/polymode/]. The algorithm solves Maxwell's equations based on the adjustable boundary condition – Fourier decomposition method (AABC_FDM) [15]. Finite differences are used in the radial direction while the Fourier decomposition method used in the angular direction helps speed up computational time, permitting faster turnaround on a high precision desktop computer.

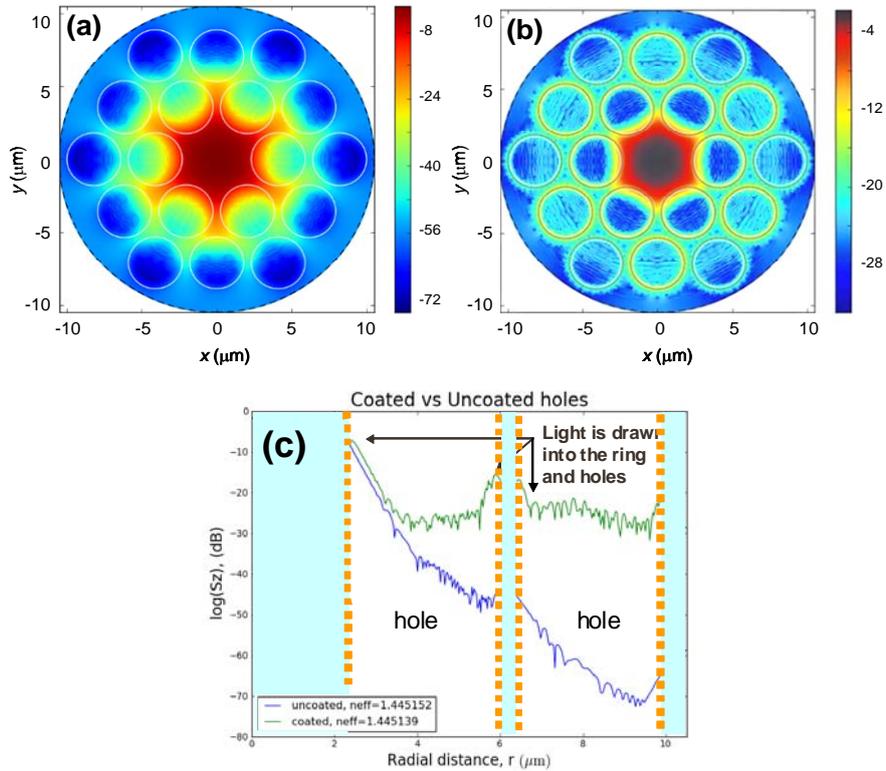

Fig. 1. Simulation of field confinement within (a) a simple 2-ring structured optical fibre; (b) the same fibre with a 155nm layer of refractive index $n$ = 2.6; and (c) cross-section of simulations showing enhanced optical localisation of light particularly near the high index surfaces (orange dashed).

Figure 1 (a) shows the numerical simulation of the fibre guided mode – the optical field intensity is plotted on log scale to exaggerate the distribution. The bulk of the optical field lies within the core defined by the first ring of holes whilst the second ring prevents any further leakage loss. Importantly, only ~ 3% of the light is in the exponentially decaying evanescent field inside the holes. Because of the low refractive index contrast between silica and air ($\Delta n$ ~ 0.45) no edge localisation of the optical field is observed. Of that evanescent light, ~ 50% is within the first (100-130) nm of the hole, signifying ~ 1.5% overlap with a nanolayer thickness $d$ ~ (100-150) nm. Despite long fibre interaction lengths, this represents an inefficient design for most sensors – multiplexed sensors, for example, would benefit from higher interaction efficiencies. In figure 1 (b) we consider the addition of a high index layer deposited onto the holes. The layer is made up of $TiO_2$ oxide in its rutile form since the fabrication of these is generally well known. After uniform film self-assembly of the crystals, this defines the layer index ($n$ = 2.6) and thickness ($d$ ~ 150nm), chosen to ensure reasonable amount of light is drawn into the holes. There are a number of interesting features observed in figure 1(b) including strong localization of the optical field both within the layer and within the surface region of the inner holes. This field appears uniform at the edges of the layers, indicative of ring resonator modes being established. The resonance modes on the channel surfaces, in particular, show the fine structure characteristic of a standing whispering gallery mode [16] that can potentially further enhance the interactions with material on these surfaces through resonance – a similar approach has been used to enhance florescence based biosensing using cylindrical cavities [17]. To illustrate more clearly the optical localization observed, especially away from the core, a cross-section of the simulated optical field is shown in figure 1 (c). There is between 2-3 dB higher signal, indicating twice as much light at the inner surface of the channel closest to the core, with much higher interrogation on the other side of the holes and in ring 2 – in total, at least an order of magnitude increase in sensitivity may be anticipated using this approach.

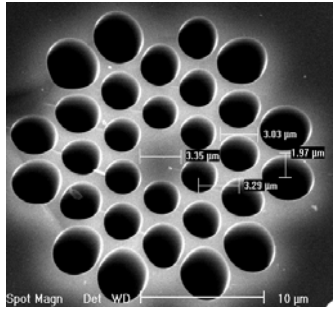

Fig. 2. SEM image of the core cross-section of a structured optical fibre with 3 rings of holes. Optical guidance is dominated by the two inner rings.

## 3. Synthesis and deposition of $TiO_2$ layers within a structured optical fiber

### 3.1 Structured optical fiber

For the actual experimental work, the structured optical fibre shown in the scanning electron microscope (SEM) image of Figure 2 was used. This fibre was assembled from Heraeus F300 silica capillaries and tubes, fused together on a lathe before being draw down to a diameter of ~125 μm using two draws on an optical fibre draw tower at ~1800°C. Custom pressure control kept the holes open both on the lathe and on the tower. Introducing a slight aperiodicity in the structure by having the holes increase in size outwards, is thought to account for the low measured propagation loss <6dB/km despite having only 3 rings [18].

### 3.2 $TiO_2$ Synthesis

A novel approach to the incorporation of $TiO_2$ layers was carried out after fibre fabrication. It involves the self-assembly of $TiO_2$ nanoparticles into low scattering films through van der Waals forces and crystal sheet formation – for this uniform size nanoparticles were required. Those obtained commercially were found to vary substantially so these were fabricated directly by hydrolysis of small quantities of titanium isopropoxide, $Ti\{OCH(CH_3)_2\}_4$, suspended in isopropanol and added to ethanol [19]. By avoiding an additional hydrothermal reaction, larger particles, chosen to ensure that there was a sufficiently thick layer to generate optical localisation of the evanescent field within the holes, could be obtained. Dynamic light scattering (DLS) measurements and transmission electronic microscopy (TEM) images confirmed that uniform $TiO_2$ nanoparticles on the order of $(155 \pm 20)$ nm were obtained. Figure 3(a) shows a monoclinic crystal configuration consistent with a partially hydrolysed form akin to so-called $TiO_2$ (B) [20] arising principally by avoiding any thermal processing.

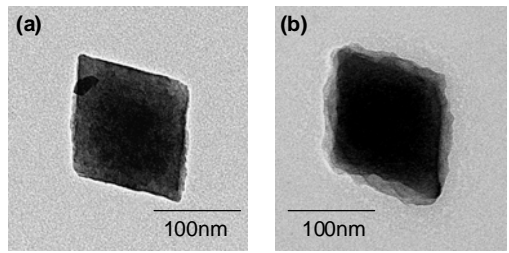

Fig. 3. TEM images of (a) crystal of $TiO_2$ showing evidence of a monoclinic unit cell and (b) similar crystal coated with TCPP.

This tends to change to tetragonal anastase and, through heating, to rutile. TiO$_2$ (B) can form thin sheet layers so under this assumption, despite a slightly lower effective refractive index (n ~ 2.5), the solution containing these monoclinic crystals was selected for insertion into the structured optical fibres.

*3.3 Spectroscopic Probe*

The spectroscopic probe chosen for these experiments was 5,10,15,20-tetra(4-carboxyphenyl)porphyrin (TCPP), where the carboxylic groups should attach well to the TiO$_2$. For all preparation conditions, the solvent employed was ethanol (EtOH) and the porphyrin concentration in EtOH was [TCPP] = 1.5 x 10$^{-3}$ M. This was reduced to [TCPP] ~7.5 x 10$^{-4}$ M after dilution and mixing with the TiO$_2$/EtOH solution (TCPP: TiO$_2$ = 1:1). After stirring, particles were filtered and examined under TEM – figure 3(b) shows evidence of attached porphyrin on a single crystal. Interestingly, as well as precipitating on the crystals, the titania generally catalysed precipitation of the porphyrins out of solution, observed both as brown aggregates which upon precipitation led to increased transparency of the solution. To verify attachment, a corresponding red-shift, $\Delta\lambda$ ~ (3-10) nm, is observed for both Soret B and Q bands in the spectra for various concentrations, shown in figure 4. The shift to longer wavelengths is consistent with J aggregation (side by side alignment) [21] rather than H aggregation (face to face) which is observed on much smaller sized nanoparticles <40 nm [22]. The general broadening of the bands reflects the size distribution of the particles.

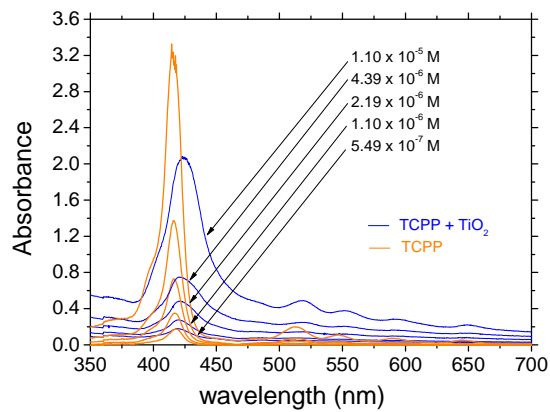

Fig. 4. UV-VIS spectra of TCPP porphyrin and TCPP porphyrin coated TiO$_2$ particles in ethanol.

*3.4 Proposed Experiments*

The method chosen to make a high index layer within the air channels of a structured optical fibre is one based on TiO$_2$ (B) forming sheets [20] through the interaction between nanoparticles the extent of which is assisted by intermolecular forces such as van der Waals forces. A reasonable assumption is that if the quality of film formation is poor, then given that these particles are commensurate in dimension with typical probe wavelengths (10-40% of visible and near IR wavelengths), substantial Mie scattering should occur, translating to readily detected propagation losses. Good film formation, on the other hand, should show little increase in loss, other than coupling losses with input and output fibres. Therefore, there are three key experiments to verify the proposed model of enhancement in our structured optical fibre. The structured fibre with the higher index layer coated with porphyrin should show much greater signal sensitivity and detection than the fibre sample with no layer but with channels coated similarly with porphyrin. Potentially, failed film formation will lead to large scattering that will undermine such a result. Therefore, a second sample with porphyrins mixed onto the TiO$_2$ prior to insertion should act as a reference from which the absolute amount of scatter may be determined, since the attached porphyrin will prevent extended $TiO_2$ layers from forming. Thus the experiments involve the separate optical interrogation of the following:

*(1)* 5,10,15,20-tetra(4-carboxyphenyl)porphyrin (TCPP) only filled fibre;
*(2)* TCPP coated $TiO_2$ particles obtained after mixing prior to insertion into fibre. In this case, the porphyrins binding to the surface will prevent film formation leading to large scattering losses; and
*(3)* TCPP and $TiO_2$ particles inserted into fibre without mixing (no coating). Film formation becomes possible leading to lower scattering losses.

## 4. Experiments

In practice, the samples are inserted under air pressure into the structured optical fibre (~3 atm) until solution is detected upon exit. A schematic of the optical interrogation configuration used is shown in figure 5. A broadband white light source (Oriel Hg-Xe lamp) is coupled into standard telecommunications fibre before being coupled into the optical fibre under test. The output is collected with another telecommunications fibre which then couples into an optical spectrum analyser (OSA – Ando AQ6315A).

The experimental results are shown in Figure 6. In the first experiment *(1)*, where the TCPP is inserted directly into the fibre with only a few per cent field in the evanescent tails within the channels, detection of the TCPP porphyrin is minimal – some of the Soret B band may be seen but otherwise there is no unambiguous detection of the weaker Q bands. In the second experiment *(2)*, where the TCPP is mixed with the nanoparticles prior to insertion, the B band is clearly picked up and evidence of the Q bands is observed. This increase in detection is clearly related to more optical field overlapping with in the layer of $TiO_2$ which contains the porphyrin. However, this has come at the expense of a large increase in background loss, consistent with Mie scattering, as expected since the attached porphyrin prevents sheet formation. In the final experiment, where the $TiO_2$ is introduced prior to the addition of the porphyrin, the largest signal detection is achieved. The signal is readily observed and the background loss has not increased substantially consistent with uniform layer formation. In contrast to the previous experiments, the Soret absorption bands are all detected clearly and unambiguously and the B band signal is stronger than that of experiment *(2)* despite detection residing mainly in the evanescent regime. This is consistent with the simulated optical localisation at the edges which is higher than within the high index regions (as shown in figure 1).

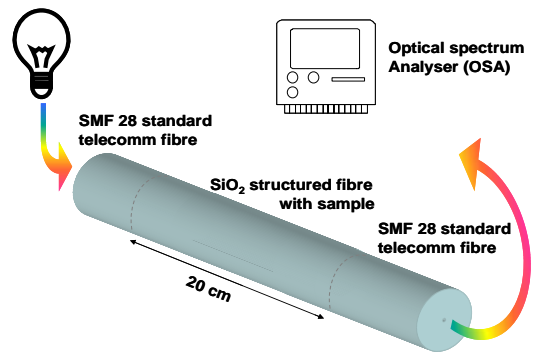

Fig. 5. Schematic of the optical interrogation setup. The spectrum within the sample fibre under test is collected using a broadband Hg-Xe white light source and optical spectrum analyser (OSA).

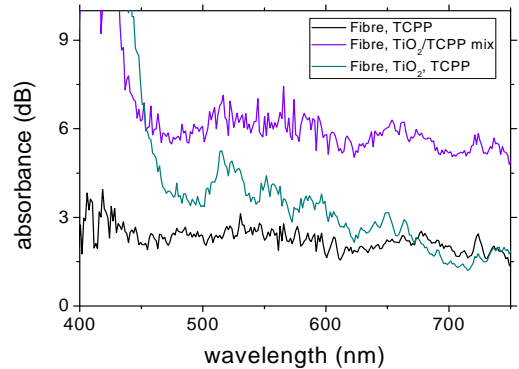

Fig. 6. Transmission spectrum of 3 structured fibre samples: (black) – with TCPP only; (purple) – with TCPP mixed with $TiO_2$ nanoparticles; and (green) with $TiO_2$ layer and TCPP. Details in the text.

## 5. Conclusion

We have proposed and demonstrated a novel approach to exploiting enhanced optical detection at low-loss structured waveguide interfaces, in this case by using a self-assembled $TiO_2$ layer within the channels of a silica photonic crystal fibre. The high index contrast between $TiO_2$ and air leads to high localisation of the evanescent field, arising from optical impedance matching at the boundary and the establishment of whispering gallery modes that resonantly enhance detection. The technique provides more than an order of magnitude greater sensitivity with minimal additional loss with standard telecommunications grade fibre coupled at input and output ends. This ensures remote optical interrogation remains feasible. More generally, the principles outlined here are not confined to the silica system and can be varied to suit many other applications, research and components technologies, including 2 and 3-D structured waveguides for example, which may require different material hosts.

**Acknowledgments**

This work was funded by several projects: Australian Research Council (ARC) Discovery Projects (DP0770692, DP0879465), and Department of Industry, Innovation, Science and Research (DISSR) International Science Linkage (CG130013). W. Padden acknowledges funding from an ARC Linkage Project (LP 0990871).